\documentclass[journal=jacsat,manuscript=article]{achemso}

\usepackage[version=3]{mhchem} 

\usepackage{float}

\usepackage{xcolor}

\newcommand{\bw}[1]{{\color{black}#1}}

\author{Milad Nourbakhsh}
\affiliation[University of Oklahoma]
{School of Electrical and Computer Engineering, University of Oklahoma, Norman, OK, 73019, USA}
\alsoaffiliation{Center for Quantum Research and Technology, University of Oklahoma, Norman, OK, 73019, USA}

\author{Kiernan E. Arledge}
\affiliation[University of Oklahoma]
{Homer L. Dodge Department of Physics and Astronomy, University of Oklahoma, Norman, OK, 73019, USA}
\alsoaffiliation{Center for Quantum Research and Technology, University of Oklahoma, Norman, OK, 73019, USA}

\author{Vincent R. Whiteside}
\affiliation[University of Oklahoma]
{Department of Electrical Engineering, University at Buffallo SUNY, Buffallo, NY 14210, USA}

\author{Jiangang Ma}
\affiliation [Northeast Normal University]
{Center for Advanced Optoelectronic Functional Materials Research, Northeast Normal University, Changchun, Jilin 130024, China }

\author{Joseph G. Tischler}
\affiliation[University of Oklahoma]
{Homer L. Dodge Department of Physics and Astronomy, University of Oklahoma, Norman, OK, 73019, USA}
\alsoaffiliation{Center for Quantum Research and Technology, University of Oklahoma, Norman, OK, 73019, USA}
\email{Tischler@ou.edu}

\author{Binbin Weng}
\affiliation[University of Oklahoma]
{School of Electrical and Computer Engineering, University of Oklahoma, Norman, OK, 73019, USA}
\alsoaffiliation{Center for Quantum Research and Technology, University of Oklahoma, Norman, OK, 73019, USA}
\email{Binbinweng@ou.edu}
\title[]
  {Phonon Polaritons and Epsilon Near Zero Modes in Sapphire Nanostructures}

\abbreviations{SPhP, RB, HVPhP, ENZ,FTIR, RS, NCA, TO, LO}

\begin{document}


\bw{

\begin{abstract}

Surface phonon polaritons (SPhPs) are promising candidates for enhanced light--matter interactions due to their efficient and low-loss light confinement features. In this work, we present unique light-matter interactions in saphhire  ($\alpha$-$\ce{Al2O3}$) within its Reststrahlen bands (RBs) across the long-wave infrared (LWIR) spectrum ($\omega = 385$--$1050~\mathrm{cm}^{-1}$). Particularly, we investigated the nanocone-patterned $\ce{Al2O3}$ resonator array, with specific attention to its in-plane and out-of-plane permittivity components. Through Fourier transform infrared spectroscopy measurement and full-wave photonic simulations, we identified a range of optical excitations in the RBs, including three SPhPs, two hyperbolic volume phonon polaritons (HVPhPs), and one epsilon-near-zero (ENZ) mode. The depth-resolved confocal Raman spectroscopy revealed strongly enhanced Raman signals on the nanostructured surface, suggesting the mode coupling between phonons and phonon-polaritons, which was further confirmed by the finite element modeling of polarizability. This exploratory study provides in-depth insights into the dynamics of LWIR phonon polaritons and ENZ modes in the nanostructured sapphire, indicating its great potential for innovative nanophotonic applications.

\end{abstract}

\section{1. Introduction}

Phonon polaritons (PhPs) are collective electromagnetic modes that arise from the coupling of photons with lattice vibrations in ionic crystals \cite{caldwell2015low}. Since phonons possess significantly longer scattering lifetimes than electrons, surface phonon polaritons (SPhPs) experience lower loss compared to surface plasmon polaritons (SPPs) \cite{caldwell2015low}. Nanoscale structures which generate SPhPs have attracted considerable attention due to their ability to manipulate light at sub-wavelength scales within Reststrahlen bands (RBs) where the material exhibits negative permittivity\cite{caldwell2015low}, and extensive research has been conducted on SPhPs in nanostructures fabricated from isotropic polar dielectrics, such as silicon carbide (SiC). These include localized SPhP from SiC nanopillar antenna arrays \cite{caldwell2013low}, high-order, multipolar SPhP from SiC  rectangular pillars\cite{ellis2016aspect}, and collective SPhP in a complex unit cells comprising SiC nanopillar subarrays\cite{lu2021collective}. The ability to harness enhanced light-matter interactions from SPhPs enables a range of applications, including sensing \cite{ellis2016aspect}, surface-enhanced spectroscopy \cite{neuner2010midinfrared,tseng2020dielectric}, and heat dissipation\cite{ghashami2018precision}.

Among ionic crystals, natural hyperbolic materials (HMs) have drawn significant attention because of their unique optical properties. These anisotropic materials exhibit both metallic-like ($\mathrm{Re}[\varepsilon] < 0$) and dielectric-like ($\mathrm{Re}[\varepsilon] > 0$) behavior along different orthogonal axes within hyperbolic RBs, facilitating access to high optical density of states and offering control over optical modes with hyperbolic dispersion properties\cite{mukhopadhyay2021natural}. Unlike SPhP modes, HMs support hyperbolic volume phonon polariton (HVPhP) modes, which lead to much larger density of states and stronger field localization than SPhPs. 

HVPhP modes have been studied in hexagonal boron nitride (hBN) nanocones\cite{caldwell2014sub, giles2016imaging}, $\alpha-\ce{Mo2O3}$ nanoribbons\cite{huang2023plane}, and calcite ($\ce{Ca2CO3}$) nanopillars\cite{breslin2021hyperbolic}. The highly directional nature of HVPhP excitations holds promise for potential applications in areas such as super-resolution imaging\cite{liu2007far} and nanolithography\cite{xiong2009simple}. Besides, sapphire ($\alpha-\ce{Al2O3}$) is another naturally occurring hyperbolic material that exhibits multiple RBs within the long-wave infrared (LWIR) spectrum. These RBs feature spectral regions that exhibit both metallic and hyperbolic optical properties, driven by the material’s anisotropic permittivity property. Although the optical properties and phonon modes of sapphire have been determined from visible to LWIR regions\cite{schubert2000infrared,stokey2022infrared}, investigations of various RBs and PhP modes in nanostructured sapphire have still been lacking.

Thus, here we present our systematic study on unique light-matter interaction phenomenon in nanostructured $\ce{Al2O3}$ - particularly of the nanocone (NC) formation across the LWIR spectrum. The work involves of experimental optical characterizations using Fourier Transform IR spectroscopy and confocal Raman scattering (RS), and theoretical study using full-wave electromagnetic simulation, to respectively probe and explain the mode resonant and phonon interaction behaviors from the $\ce{Al2O3}$ NC structure. Through which, it will present you an in depth understanding of PhPs and ENZ modes in $\ce{Al2O3}$ nanostructures that was lacking at this time. Understanding several RBs in $\ce{Al2O3}$ will also provide insight into the physics of emerging phononic-induced magnetization switching systems\cite{davies2024phononic}. 

\section{2. Optical properties of c-plane bulk \ce{Al2O3}}

Figure~\ref{fig1}a shows the crystallographic unit cell of $\alpha-\ce{Al2O3}$ aligned with the laboratory coordinates. The crystal exhibits an identical atomic structure perpendicular to the $c$-axis in both the $x$ and $y$ directions (Fig. \ref{fig1}a, top right panel), while the atomic arrangement differs along the $z$ direction, parallel to the $c$-axis (Fig. \ref{fig1}a, bottom right panel), resulting in anisotropic optical properties of the crystal. 

\begin{figure}
  \centering
\includegraphics[width=6 in,height=5.5 in]{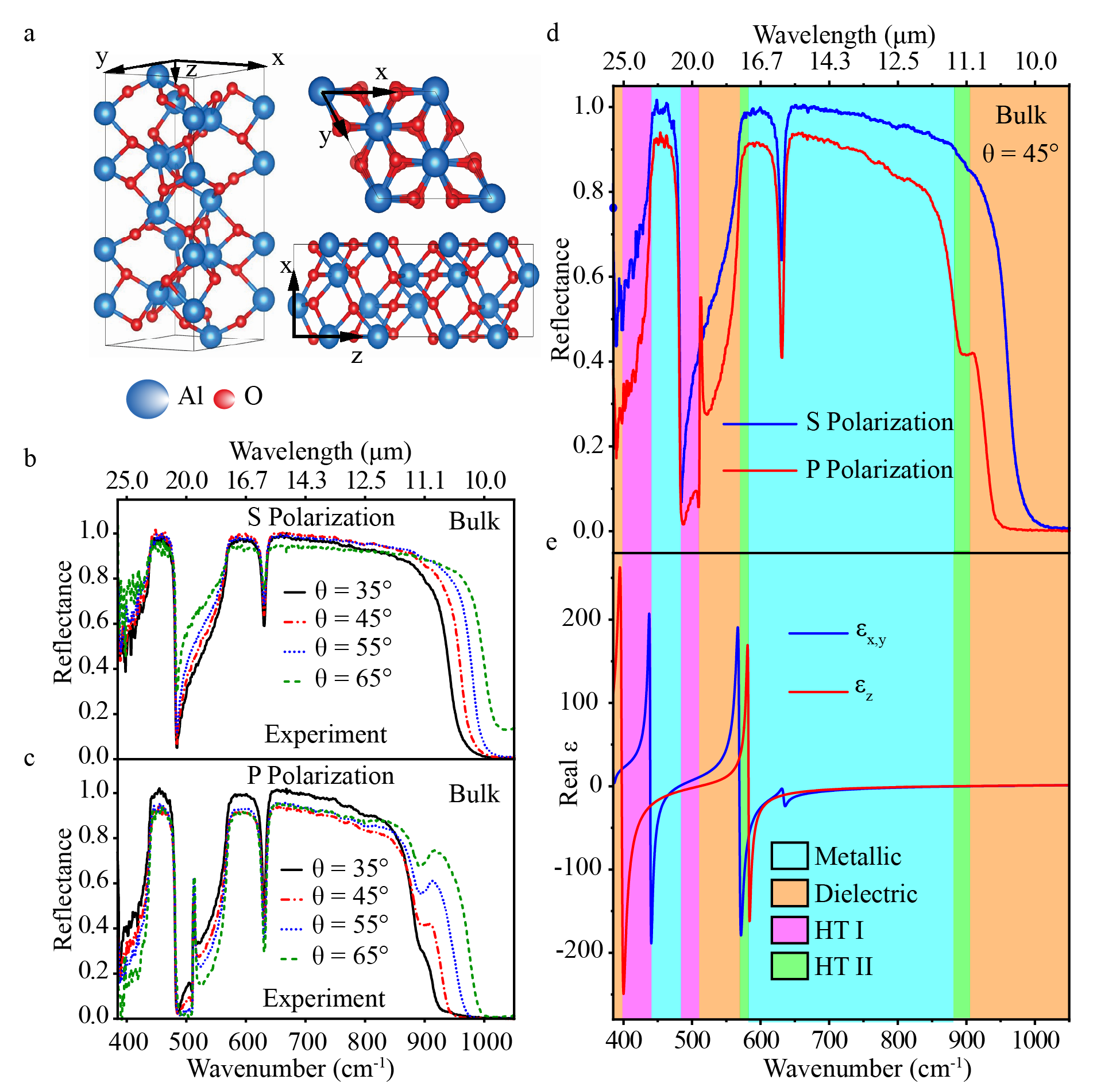}
  \caption{(a) Schematic of the crystallographic unit cells for $\alpha$-$\ce{Al2O3}$. The left panel shows the 3D atomic structure, while the right panels show $xy$ and $xz$ orientations. The black and red circles indicate aluminum and oxygen atoms, respectively. (b) Angle-dependent reflectance measurements for $s$-polarized excitation (polarization is perpendicular to the optic axis) and (c) for $p$-polarized excitation (polarization is parallel to the optic axis) for the bulk material. (d) Reflectance measurements for $s$ and $p$-polarized excitations for the bulk material with $\theta = 45^\circ$. (e) The real part of permittivity components for (001) sapphire ranging from 385~cm$^{-1}$ to 1050~cm$^{-1}$. The blue curve shows permittivity in $x$  and $y$ directions (perpendicular to the optic axis), and the red curve indicates the permittivity along $z$ direction (parallel to the optic axis). Color coding indicates various RBs of the material which are metallic (cyan), hyperbolic type $\mathrm{I}$ (magenta), and type $\mathrm{II}$ (green).
}
  \label{fig1}
\end{figure}

This anisotropy is demonstrated in figures~\ref{fig1}b and c, which present the IR reflection spectroscopy (See Methods Section~II) for both $s$ and $p$-polarized light at various angles of incidence (Fig. S1). Significant variations in reflectivity were observed for both polarizations, particularly at the high-energy edges of the reflection bands, as the angle of incidence changed. This is primarily attributed to the critical condition for the material’s total reflection (further details are provided in SI)\cite{schubert2000infrared}. 

Group theory predicts five pairs of IR-active optical phonons within the studied spectral range: three with $E_u$ symmetry, which are characterized by dipole oscillations perpendicular to the optic axis ($\mathbf{E} \perp c$), and two with $A_{2u}$ symmetry, where dipole oscillations occur parallel to the optic axis ($\mathbf{E} \parallel c$)\cite{schubert2000infrared, stokey2022infrared}. The three reflection bands observed in the $s$-polarized reflectivity (Fig.~\ref{fig1}b) are each bounded with $E_u$ symmetry transverse optic (TO$_{E_u}$) and longitudinal optic (LO$_{E_u}$) phonons. In contrast, $p$-polarized reflectivity comprises five reflection bands arising from the combination of three $E_u$ and two $A_{2u}$ phonon pairs, as shown in Fig.~\ref{fig1}c\cite{schubert2000infrared}. In contrast to many polar dielectric systems which have been studied in the IR spectrum, such as SiC with a metallic RB\cite{caldwell2013low, ellis2016aspect, lu2021collective} and hBN\cite{caldwell2014sub, giles2016imaging}, which exhibits two hyperbolic RBs (i.e., spectral regions where $\mathrm{Re}[\varepsilon_{xx,yy}] \cdot \mathrm{Re}[\varepsilon_{zz}] < 0$), $\ce{Al2O3}$ is a naturally hyperbolic material featuring both metallic and hyperbolic windows within its RBs. This unique characteristic allows for the excitations of both SPhP and HVPhP within the IR spectrum. 

The color coding in Figs. \ref{fig1}d and \ref{fig1}e highlights distinct spectral regions of the material, based on the sign of the real parts of dielectric function for both axes, as shown in Fig. \ref{fig1}e. Each color corresponds to a specific spectral range based on the sign of the in-plane ($\varepsilon_{xx,yy}$) and out-of-plane ($\varepsilon_{zz}$) permittivity components, resulting in intervals with distinct optical behaviors, including metallic, dielectric, and hyperbolic properties. As shown in Fig. \ref{fig1}e, the magenta regions correspond to type~I hyperbolicity (HT~I), where the material exhibits metallic-like behavior along the $z$~direction with negative $\mathrm{Re}[\varepsilon_{zz}]$, and dielectric-like behavior along other directions with positive $\mathrm{Re}[\varepsilon_{xx,yy}]$. Conversely, the green-colored regions, corresponding to type~II hyperbolicity (HT~II) show the opposite behavior: metallic-like properties are displayed in-plane and dielectric-like properties along the $z$~direction. Additionally, the cyan regions correspond to metallic behavior, where all permittivity components are negative resulting in total reflection bands, as seen in Fig. \ref{fig1}d. In contrast, the orange regions indicate the material behaves as an anisotropic dielectric, with positive permittivity components in all directions (The exact ranges of various spectral regions are presented in Table S1). Understanding the various bands of $\ce{Al2O3}$ is crucial for advancing research on heterostructures incorporating sapphire, including novel phonon-induced magnetization switching systems\cite{davies2024phononic}. 

\section{3. Light-matter interactions from nanostructured \ce{Al2O3}}

PhPs are regarded as quasiparticles with high in-plane momentum relative to that of free-space photons, and therefore cannot be directly excited in the bulk of a material by far-field radiation\cite{caldwell2015low}. Instead, surface structuring can overcome this momentum mismatch, such as using a nanocone array (NCA) structure. Figs. \ref{fig2}a and \ref{fig2}b show both experimentally measured and numerically calculated reflectivity for the NCA sapphire, along with the experimental reflectivity of the bulk material, for both $s$ and $p-$polarized excitations with oblique incidence of $\theta = 45^\circ$. In both experimental spectra, we observed regions of significantly reduced reflection, attributed to the excitation of various polariton modes. A comprehensive study of the NCA optical response, including measurements of the IR reflectivity at various polarizations and oblique angles of incidence has also been conducted, verifying the robust presence of the observed reflection dips (see SI, Fig. S3). 

Full-wave simulations were conducted for the NCA \ce{Al2O3} using COMSOL Multiphysics. The calculated reflectivity plots are shown as dashed lines in Figs. \ref{fig2}a, \ref{fig2}b, and \ref{fig2}d (see SI). To model the optical response, the structure was characterized by Scanning Electron Microscopy (SEM) (see Methods section I). Figure \ref{fig2}c shows an SEM image with the left panel showing the side view and the right panel showing the top view of the structure. The NC's hexagonal array has a periodicity of approximately $p = 3.2~\mu\text{m}$, a heigh of $h = 2~\mu\text{m}$,  and a basal diameter of $d = 3~\mu\text{m}$. The green rectangles in the right panel indicate periodic unit cells and boundary conditions (BC) used in modeling. As evident from the reflection spectra in Figs. \ref{fig2}a and \ref{fig2}b, three SPhP modes are excited by the NCA structure, indicated by reflection dips superimposed on the metallic RBs (Fig. \ref{fig1}d) \cite{caldwell2015low}. 

\begin{figure} 
  \centering
\includegraphics[width=6 in,height=5.5 in]{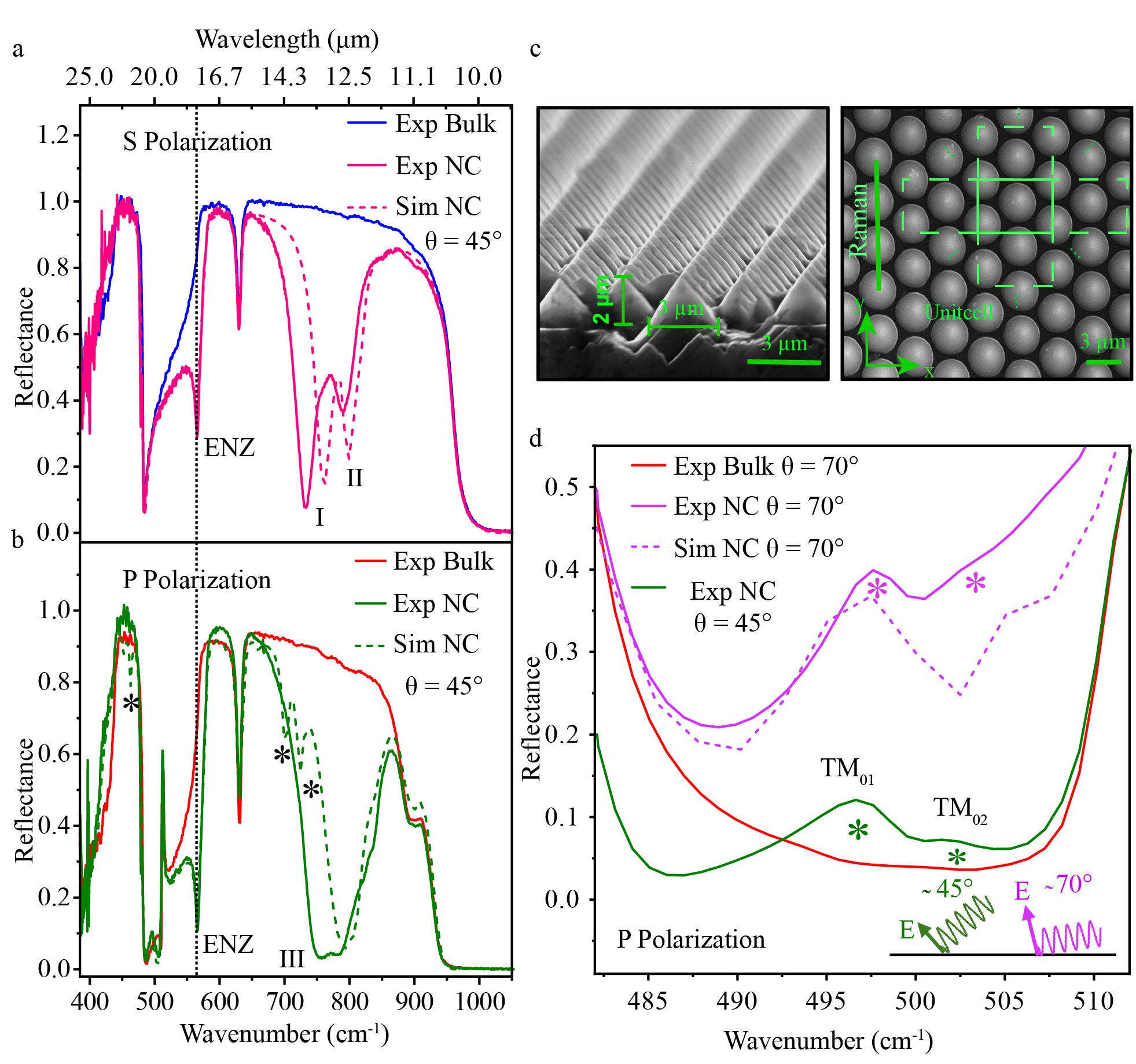}
  \caption{\textbf{Optical properties of nanocone-structured sapphire}(a) Reflection spectra for the bulk and nanocone-structured material for the oblique $s$-polarized and (b) $p$-polarized incidence with angle of $\theta = 45^\circ$. The solid curves show reflection measurements for the bulk and nanostructured samples, and the dashed curves represent the simulation results obtained from modeling of the nanocone arrays using COMSOL Multiphysics. (c) The SEM images indicate the side view (left panel) and the top view (right panel) of the nanocone-structured sample. The nanostructure dimensions and the periodic rectangular unitcell used in the simulation shown on the images. (d) Reflection spectra for the bulk and nanostructured samples for $p$-polarized incidence with angles of $\theta = 45^\circ$ and $\theta = 70^\circ$  within the spectral range of 482\,cm$^{-1}$ to 511\,cm$^{-1}$ where the material exhibits hyperbolic type~I behavior. The solid curves represent reflection measurements for the bulk and nanostructured samples, and the dashed curves demonstrate simulation results.}
  \label{fig2}
\end{figure}

For the $s-$polarized light, the experimental data (solid lines) reveal two SPhP modes, denoted~I and~II, at wavenumbers $\omega = 734~\text{cm}^{-1}$ and $\omega = 790~\text{cm}^{-1}$ (wavelengths $\lambda = 13.6~\mu\text{m}$ and $\lambda = 12.6~\mu\text{m}$, respectively). For $p-$polarized excitation, a broad SPhP resonance, labeled~III, is observed at $\omega = 772~\text{cm}^{-1}$ (wavelength $\lambda = 13~\mu\text{m}$). Further details on mode~III is provided in supporting information, Fig. S4. The measured and calculated data exhibit a good agreement, except for discrepancies in the SPhP resonance frequencies. These modes depend on microscopic features of the nanostructure and minor discrepancies are suggested to be attributed to small irregularities in the shape of the actual nanocone (NC) compared to the idealized NC employed in the modeling \cite{caldwell2014sub}. Additionally, three resonances are observed in the calculated reflectivity at spectral positions $\omega = 463~\text{cm}^{-1}$, $\omega = 699~\text{cm}^{-1}$, and $\omega = 724~\text{cm}^{-1}$, as indicated by asterisks in Fig. \ref{fig2}b. These resonances are absent in the measured reflectance spectrum. However, given their weaker amplitudes compared to the other measured modes, these resonances are likely below the detection threshold of our measurement systems \cite{ellis2016aspect} (Refer to SI Figs. S5g, S5h, and S5i for the E-field profiles corresponding to these resonances). 

The corresponding E-field spatial distributions in $ZY$ plane for each mode is shown in color plots in Fig. \ref{fig3}. Numerical calculations of $E_y$-field spatial distribution at normal incidence reveal that mode~I is a localized surface confinement along the face of the NC (Fig. \ref{fig3}d). Mode~II is a transverse dipole with optical fields concentrated at the apex and troughs of the NC (Fig. \ref{fig3}e). Since $s-$polarized light is perpendicular to the plane of incidence, it primarily excites transverse surface modes along the NC. Figure \ref{fig3}f shows the calculated $E_z$-field spatial distribution for mode~III at normal incidence, where it is clear that $p-$polarized light enhances the vertical confinement in the tip, sides, and troughs of the NC. In contrast, $p-$polarized light with its $E_z$ component primarily excites vertical modes in the nanostructure. The relatively broad linewidth of mode~III may be attributed to the tapered shape of the NC along the $z$~direction\cite{cui2012ultrabroadband, yang2018broadband}. 

\begin{figure} 
  \centering
\includegraphics[width=6 in,height=5.5 in]{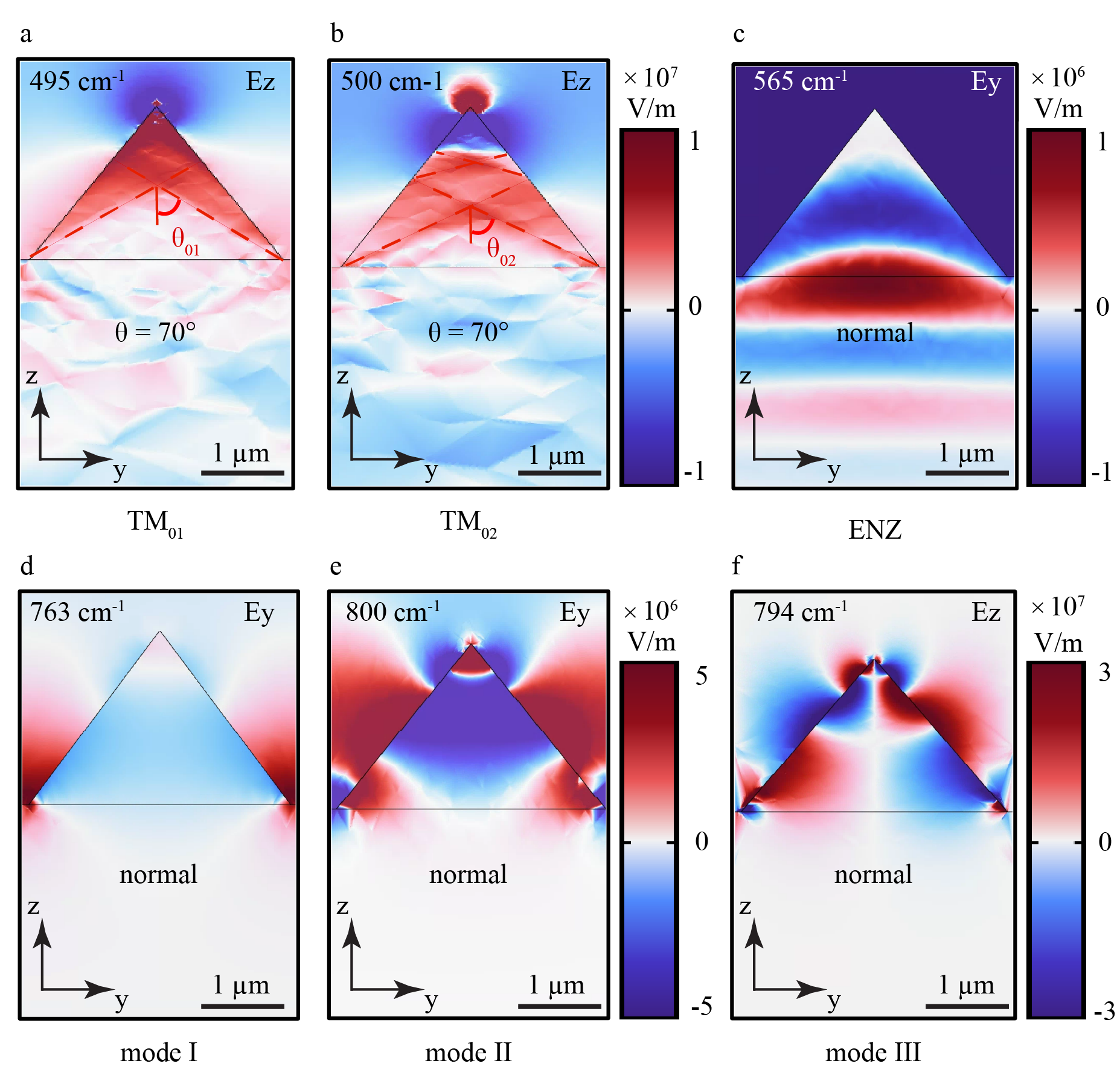}
  \caption{\textbf{E-field spatial distribution in NCA sapphire system.
} $E_z$ simulated cross-sectional plots for HvPhP at (a) 495~cm$^{-1}$, and (b) 500~cm$^{-1}$ with near grazing incidence of $\theta = 70^\circ$. The normal incidence $E_y$ simulated cross-sectional plots for (c) ENZ mode at 565~cm$^{-1}$, and for $s$-polarized SPhP resonances at (d) 763~cm$^{-1}$, and (e) 800~cm$^{-1}$. Normal incidence $E_z$ surface plot for $p$-polarized SPhP at (f) 794~cm$^{-1}$.
}
  \label{fig3}
\end{figure}

In addition to SPhPs, HVPhPs are excited in the NCA platform under $p-$polarized excitation within the HT~I spectral region from 482~cm$^{-1}$ to 511~cm$^{-1}$. As demonstrated in Fig. \ref{fig2}d, the dominant HVPhPs manifest as reflection peaks at $\omega = 497~\text{cm}^{-1}$ and $\omega = 502~\text{cm}^{-1}$ (wavelengths $\lambda = 20.1~\mu\text{m}$ and $\lambda = 19.9~\mu\text{m}$, respectively). Measurements taken at off-normal incidence ($\theta = 45^\circ$) and near-grazing incidence ($\theta = 70^\circ$), reveal that the strength of the HVPhPs is dependent on the incident angle. As the incident angle increases, the electric field along the material’s metallic direction ($E_z$) intensifies, resulting in stronger resonances in the solid magenta plot compared to the green reflectivity plot. Figures \ref{fig3}a and \ref{fig3}b illustrate calculated $E_z$-field distribution and provide further insight into the nature of the HVPhPs. The cross-hatch $E-$field profile is attributed to the propagation direction of the HVPhPs which is restricted to the angle between the field crossing lines and the optical axis ($\theta_{01}, \theta_{02} = \arctan\left( \sqrt{\varepsilon_z(\omega)}\, /\, i \sqrt{\varepsilon_{xy}(\omega)} \right)$
 define the propagation directions of the HVPhPs in Figs. \ref{fig3}a and \ref{fig3}b)\cite{caldwell2014sub}. Given that the NC platform features a spheroid-like geometry with its symmetrical axis aligned parallel to the $c$-axis, the HVPhP eigenmodes are denoted as TM$_{ml}$, where TM (transverse magnetic) indicates that the magnetic field is perpendicular to the $z$ axis \cite{caldwell2014sub}. The parameter $m$ represents the angular momentum along the $z$ axis, and $l$ corresponds the orbital index \cite{caldwell2014sub}. Since the HVPhPs are symmetric modes around the $z$ axis, $m = 0$ for both observed modes \cite{caldwell2014sub}. 
 
 Additionally, $l = 1$ characterizes the first-order mode with a single node in the $E-$field profile, while $l = 2$ denotes the second-order mode, which features two nodes in the $E-$field profile, as illustrated in Figs. \ref{fig3}a and \ref{fig3}b, respectively. The directional HVPhPs supported by the sapphire NCA platform makes it a promising candidate for potential applications such as super-resolution imaging and nanolithography\cite{liu2007far, xiong2009simple}. Furthermore, outside the HT~II spectral region (from 569~cm$^{-1}$ to 583~cm$^{-1}$), near the transverse optical phonon TO$_{E_{u_2}}$, a strong resonance observed at $\omega = 565~\text{cm}^{-1}$ ($\lambda = 17.7~\mu\text{m}$) in both $s$ and $p-$polarized reflection spectra, connected by a dashed line in Fig. \ref{fig2}a and \ref{fig2}b. This epsilon-near-zero (ENZ) state owes its existence to the strong dispersion of the material’s in-plane permittivity ($\mathrm{Re}[\varepsilon_{x,y}]$) close to the hyperbolic transition point, enabling enhanced light confinement inside the NC structures as depicted in Fig. \ref{fig3}c \cite{fomra2024nonlinear} (Refer to SI Fig.~S5 for the $E$-field profiles of the observed modes for the oblique incidence of $\theta = 45^\circ$.)

\section{4. Raman enhancement from nanostructured \ce{Al2O3}}

IR-active modes must involve a change in dipole moment, and Raman-active modes must involve a change in polarizability. In centrosymmetric materials, symmetry operations prevent the same vibrational mode from satisfying both conditions simultaneously. Since, sapphire is a centrosymmeric material, it also follows the mutual exclusion principle\cite{watson1981measurements},  meaning that all IR related vibration modes discussed in previous section are not Raman-active modes. Therefore, experimental investigation of the Raman-active modes using Raman spectroscopy can serve as a complementary approach to gain deeper insights into the unique light–matter interaction phenomena and underlying mechanisms in nanostructured \ce{Al2O3}.

\begin{figure}[H]
  \centering
\includegraphics[width=6.5 in,height=3.5 in]{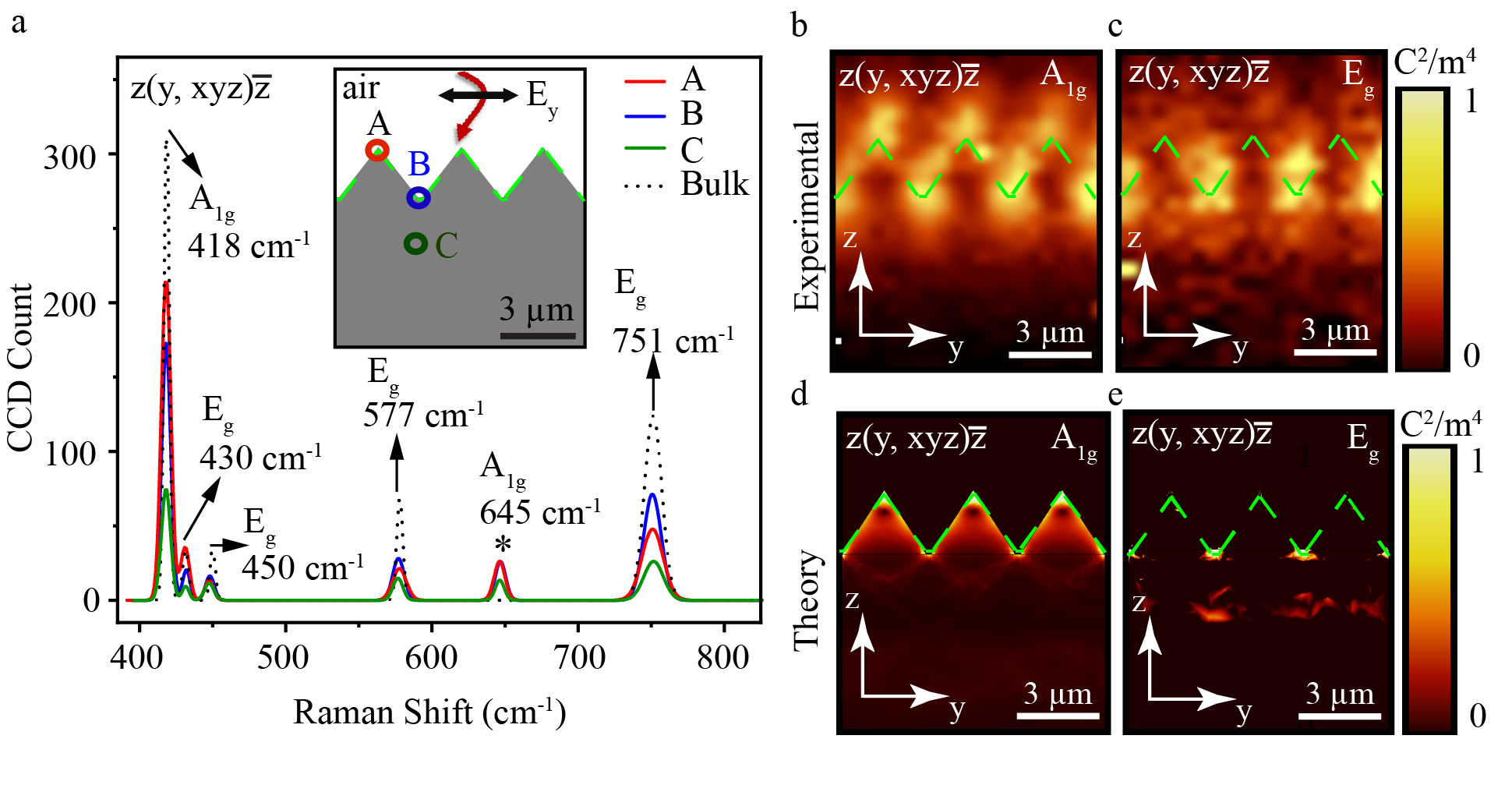}
  \caption{\textbf{Mapping PhPs in \ce{Al2O3} NCA.
} (a) Measured Raman at three different points (as shown in the inset) across the NCA indicating the spatial dependence of Raman modes. $ZY$ cross sectional view of Raman intensity obtained from experimental measurements for (b) $A_{1g}$ phonon mode at $\omega= 418~cm^{-1}$, and (c) $E_{g}$ phonon mode at $\omega= 751~cm^{-1}$  for three adjacent NCs. Theoretically simulated  Raman intensity shown in $ZY$ cut plane for (d) $A_{1g}$ and (e)  $E_{g}$ phonon modes. The dashed green lines indicate the NCs-air interface.
}
\label{fig4}
\end{figure}

The Raman-active phonons in bulk \ce{Al2O3} material have been studied before. Within the investigated range, six Raman modes have been reported for the material: two classified as A$_{1g}$ and four as E$_g$ modes\cite{watson1981measurements, wermelinger20073, zhu2011raman, talwar2020anisotropic, tao2023probing, bergeron2023probing}. To explore the spatial variation of the vibrational modes in this nanostructured \ce{Al2O3}, confocal Raman scattering (RS) measurements were performed at various points on its NC structure: the apex, the trough, and 2~$\mu$m deep into the substrate (see Methods section~III for more information about RS measurements). The results indicated by the red, blue, and green plots in Fig.~\ref{fig4}a correspond to the scan points A, B, and C, respectively. The distinct spectral responses highlight the significant spatial dependency of RS on the NC geometry. Additionally, a  Raman enhancement observed at $\omega= 645~cm^{-1}$ for NCA structure, which is absent in the bulk material, as shown with asterisk in Fig.~\ref{fig4}a. The $A_{1g}$ Raman mode appears only in the diagonal components of the Raman tensor, particularly in the $z$ direction. As a result this mode is absent in $c$-cut sapphire when RS measurement is conducted using $z(y, )\bar{z}$ geometry \cite{watson1981measurements, wermelinger20073}. However, the same mode has been observed in $a$-cut sapphire under similar measurement configuration \cite{tao2023probing}. This Raman enhancement may be attributed to the relaxation of Raman selection rules in nanostructured system leading to the activation of forbidden vibrational mode in perfect crystal \cite{sirleto2017advances, tan2019raman}. 

Additionally, two-dimensional (2D) cross-sectional Raman spatial mapping was performed by scanning three adjacent NCAs in the $ZY$ plane (the right panel in Fig.~\ref{fig2}c shows the top view of the 2D scanned plane). As depicted in Fig.~ \ref{fig4}b and \ref{fig4}c, the Raman mapping extended from 2 $\mu\text{m}$ above the NC apex to 6 $\mu\text{m}$ deep into the substrate. These Raman images visualize coupling between bulk phonons and PhP, including SPhP and HVPhP \cite{bergeron2023probing, arledge2024mapping} (SI Fig.~S6). The Raman image in Fig.~\ref{fig4}b shows that the $A_{1g}$ mode exhibits strong intensity at the apex, troughs, and within the volume of the NC, analogous to SPhP mode~II and HVPhP modes. In contrast, Fig.~\ref{fig4}c demonstrates that the $E_g$ mode intensity is stronger at the NC troughs, like the SPhP mode~I. This variation in coupling is driven by changes in local polarizability, described by Raman selection rules \cite{arledge2024mapping}. To further support our experimental data, the Raman selection rules for both $A_{1g}$ and $E_g$ modes were employed to reconstruct the field intensity within the NCAs. Raman intensity corresponding to these phonon scattering modes was theoretically simulated with COMSOL Multiphysics, with the results shown in Fig.~\ref{fig4}d and \ref{fig4}e. The calculated Raman data shows excellent agreement with the experimental findings, visualizing the coupling between phonons and PhPs in NC structure \cite{arledge2024mapping}.

\section{Conclusion}
In conclusion, our study provides a thorough analysis of the optical properties of $\ce{Al2O3}$ within the IR spectral range from $\omega = 385-1050~ 
 cm^{-1}$. We demonstrated that sapphire is a complex polar dielectric material featuring both hyperbolic and metallic RBs. Through experimental analysis and full-wave simulation, we demonstrated strong light-matter confinement bellow the diffraction limit within the sapphire NC structure, arising from excitation of PhPs and  ENZ modes. Specifically, three SPhPs, two HVPhPs and one ENZ modes observed in the studied nanostructure. Furthermore, confocal RS revealed enhanced Raman signals on the nanostructured surface and demonstrated phonon–PhPs coupling through spatial mapping. This study confirms that the sapphire NCA structure can serve as a promising platform for advanced IR nanophotonic applications and provides valuable insights into the physics of the novel remote phonon-induced magnetization switching systems, leveraging phononic properties of sapphire.

\section{Methods}

\subsection {I. Scanning electron microscopy (SEM)}
The patterned sapphire substrates were purchased from Crystal Optoelectronics Company in Zhejiang, China. The preparation process of patterned sapphire substrate starts from spin-coating a layer of photoresist on the surface of the sapphire substrate. Through the steps of pre-backing, alignment, ultraviolet light exposure, baking, development, etc., the pattern on the mask is replicated onto the photoresist. Then, using the patterned photoresist film as a mask, an etching process was performed to obtain a patterned sapphire substrate. The patterned sapphire substrate morphology was measured using the field-emission scanning electron microscope (SEM, Hitachi S-4800).

\subsection{II. Far-field measurement using Fourier Transform Infrared (FTIR) spectroscopy}
Reflectance measurements for bulk and NC sapphire samples (Figs.~ \ref{fig1}c, \ref{fig1}d, \ref{fig2}a, \ref{fig2}b, and \ref{fig2}d in the main text and Figs. S2 and S3 in the supporting information) were carried out using a Bruker INVENIO\textsuperscript{R} FTIR spectrometer. The spectrometer, equipped with a KBr beam splitter, can cover the spectral range from 0.25~$\mu$m to 25~$\mu$m, broad spectral range DLaTGS detector, and PIKE Technologies VeeMAX\textsuperscript{TM} III specular reflectance accessory. The VeeMAX\textsuperscript{TM} III reflectance accessory enables the samples to be analyzed over a range of incident angles from 30$^\circ$ to 80$^\circ$. Thorlabs' holographic wire grid polarizer, which operates over the 2-30~$\mu$m spectral range (333--5000~cm$^{-1}$), was used to illuminate the samples under various polarizations. All measurements were carried out under room temperature and ambient pressure conditions, with nitrogen gas purging to remove moisture interference. The reflection spectra were recorded in the wavenumber range of 385~cm$^{-1}$ to 1050~cm$^{-1}$ (corresponding to a wavelength range of 9~$\mu$m to 26~$\mu$m) with a scan number of 60 and a spectral resolution of 4~cm$^{-1}$.

\subsection{III. Confocal Raman spectroscopy}
Raman spectra were acquired using a WITec Alpha 300R confocal Raman microscope equipped with a 100$\times$ Zeiss EC Epiplan-Neofluar objective (0.95 NA), an ultra-high throughput spectrometer WITec UHTS 300, a charged-coupled device (CCD) detector, and a $\lambda/2$ calcite waveplate. All Raman scattering measurements were performed using $z(y, )\bar{z}$ geometry with a green laser centered at $\lambda = 532$~nm, where the $y$-polarized incident and unpolarized scattered light propagated along the $z$ axis. The Raman scattering measurements at various spatial points on the  nanostructure and the bulk material (Fig.~\ref{fig4}a in the main text) were collected at room temperature and under ambient pressure condition with accumulation and integration times of 50 and 25 seconds, respectively. Additionally, a similar setup was used to measure the $ZY$ cross-sectional spatial Raman intensity for the nanostructure (Figs.~\ref{fig4}b and~\ref{fig4}c in the main text and Figs.~S6g and S6h in the supporting information). To measure the spatial Raman intensity shown in Figs.~\ref{fig4}b and~\ref{fig4}c in the main text, a Raman depth scan was performed on three adjacent NCs across the $ZY$ plane (a green line in Fig.~\ref{fig2}c in the main text shows the top view). The scan resolution was 0.5~$\mu$m, i.e., Raman intensity was measured at every 0.5~$\mu$m interval in both the $z$ and $y$ directions across the selected plane. Each point was scanned with an integration time of 25 seconds, allowing for precise mapping of the spatial variations in both dimensions.

\begin{acknowledgement}
B.W. acknowledge partial supports by U.S. National Science Foundation (NSF) CAREER Award under grant No. 2340060, and the NSF Partnership for International Research and Education (PIRE) program under the grant No. 2230706. J.G.T. acknowledges support by U.S. NSF under Grant OISE-2230706 and Oklahoma Center for the Advancement of Science and Technology’s Research Grant No. AR21-032.  M.N. acknowledges Mathias Schubert and Megan Stocky for providing sapphire refractive index data used in the modeling for both bulk and nanostructured material. 
\end{acknowledgement}
}


\section{Supporting Information} 
The Supporting Information provides additional data relevant to the reflection bands of bulk sapphire, including Table S1, which lists the exact spectral ranges of the material's various Reststrahlen bands. Furthermore, Figures S1-S6 represent additional experimental and numerical results, including reflection spectra and sub-diffraction mode profiles in nanocone-structured sapphire.

\setcounter{figure}{0}
\renewcommand{\thefigure}{S\arabic{figure}}
\setcounter{table}{0}
\renewcommand{\thetable}{S\arabic{table}}

\subsection{Reststrahlen bands of Saphhire in the LWIR region}

To better understand the observed dynamics in sapphire reflection bands, $s$ and $p$-polarized reflectivity at off-normal incidence  $\theta = 45^\circ$, and real parts of the material’s in-plane and out-of-plane permittivities, determined by spectroscopic ellipsometry (SE)\cite{schubert2000infrared}, are shown in Fig.~1(d) and 1(e) in the main text, respectively. The total reflection occurs at incident angles greater than the critical angle where the condition $\varepsilon(\omega) < \varepsilon(\omega_{\text{cut-off}})$ is satisfied. Here, $\omega_{\text{cut-off}}$ represents cut-off frequency and is defined as $\omega_{\text{cut-off}} = \sin^2\theta$, where the material’s reflection band drops significantly, and $\theta$ is the the angle of incidence \cite{schubert2000infrared}. The cut-off frequency undergoes  blueshift as the angle of incidence increases, as shown in Figs.~ 1(b) and 1(c) in the main text. For example, $s$-polarized light with incident angles of $\theta = 45^\circ$ and $\theta = 65^\circ$ leads to a significant drop in reflectivity at $\omega_{\text{cut-off}} \approx 962~\text{cm}^{-1}$ and $\omega_{\text{cut-off}} \approx 1006.5~\text{cm}^{-1}$, respectively. 

Additionally, because of the material's anisotropic feature and  gradual variation in permittivity components $\omega_{\text{cut-off},z} < \omega_{\text{cut-off},xy}$, resulting in the $p$-polarized high-energy reflection band drops at lower energies than the $s$-polarized reflectivity \cite{schubert2000infrared} (Fig.~1e).  Furthermore, a sharp narrow band at $\omega \approx 511~\text{cm}^{-1}$ corresponds to excitation of $\mathrm{LO}_{A_{2u_1}}$ phonon driven by the $z$ component electric field in $p$-polarized light~\cite{schubert2000infrared}. Here, the significant optical loss of the $\mathrm{LO}_{A_{2u_1}}$ phonon inhibits near-unity reflection~\cite{schubert2000infrared}. As shown in Fig.1(c), the sharpness of this peak is influenced by the angle of incidence. As the incident angle increases, the $z$ component of the $p$-polarized electric field becomes more prominent, leading to stronger interaction between the $\mathrm{LO}_{A_{2u_1}}$ phonon and light, thereby enhancing reflectivity. Moreover, because of the overlap between $A_{2u_2}$ and $E_{u_3}$ frequency ranges, the reflection condition is not satisfied, and reflection loss is observed at $\omega \approx 893~\text{cm}^{-1}$ in $p$-polarized spectra~\cite{schubert2000infrared}. Furthermore, the spectral ranges corresponding to the material's various Reststrahlen bands (RBs) are detailed in Table~\ref{tblS1}.
\begin{table}[htbp]
  \centering
  \caption{Spectral range of sapphire various Reststrahlen bands}
  \label{tblS1}
  \begin{tabular}{ccc} 
    \hline
    Optical Response & Wavenumber ($\text{cm}^{-1}$) & Wavelength ($\mu\text{m}$) \\
    \hline
    HT~I & 397--439 & 22.8 - 25.2 \\
    Metallic          & 439--482 & 20.8 - 22.8 \\
    HT~I & 482--511 & 19.6 - 20.8 \\
    Dielectric        & 511--569 & 17.6 - 19.6 \\
     HT~II & 569--583 & 17.2 - 17.6 \\
    Metallic          & 583--881 & 11.4 - 17.2 \\
     HT~II & 881--906 & 11 - 11.4 \\
    Dielectric        & 906 - 1050 & 9.5 - 11 \\
    \hline
  \end{tabular}
\end{table}

\subsection{Numerical modeling comparison with experiments}
\begin{figure}
  \centering
\includegraphics[width=6.5 in, height=3.5 in]{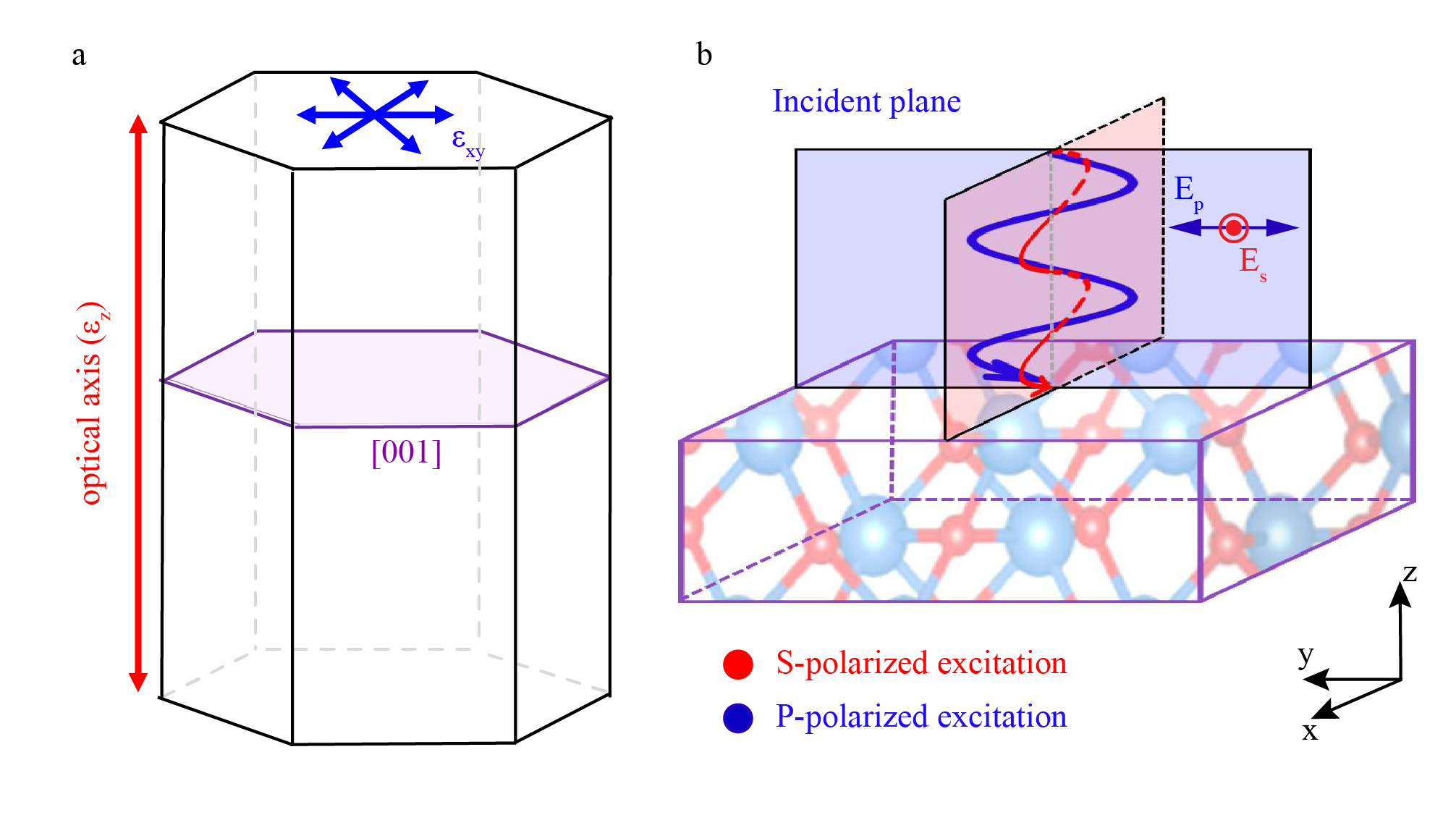}
  \caption{(a) A (001) sample cut from a hexagonal crystal unit cell, with the optic axis ($c$-axis) oriented perpendicular to the sample’s surface. (b) Polarized components of a normal incident light beam. Here, the $p$-polarized light is in the $YZ$ plane, while the $s$-polarized light is positioned in $XZ$ plane, perpendicular to the plane of incidence.
}
  \label{FigS1}
\end{figure}

\begin{figure}[H]
  \centering
\includegraphics[width=6.5 in, height=3.5 in]{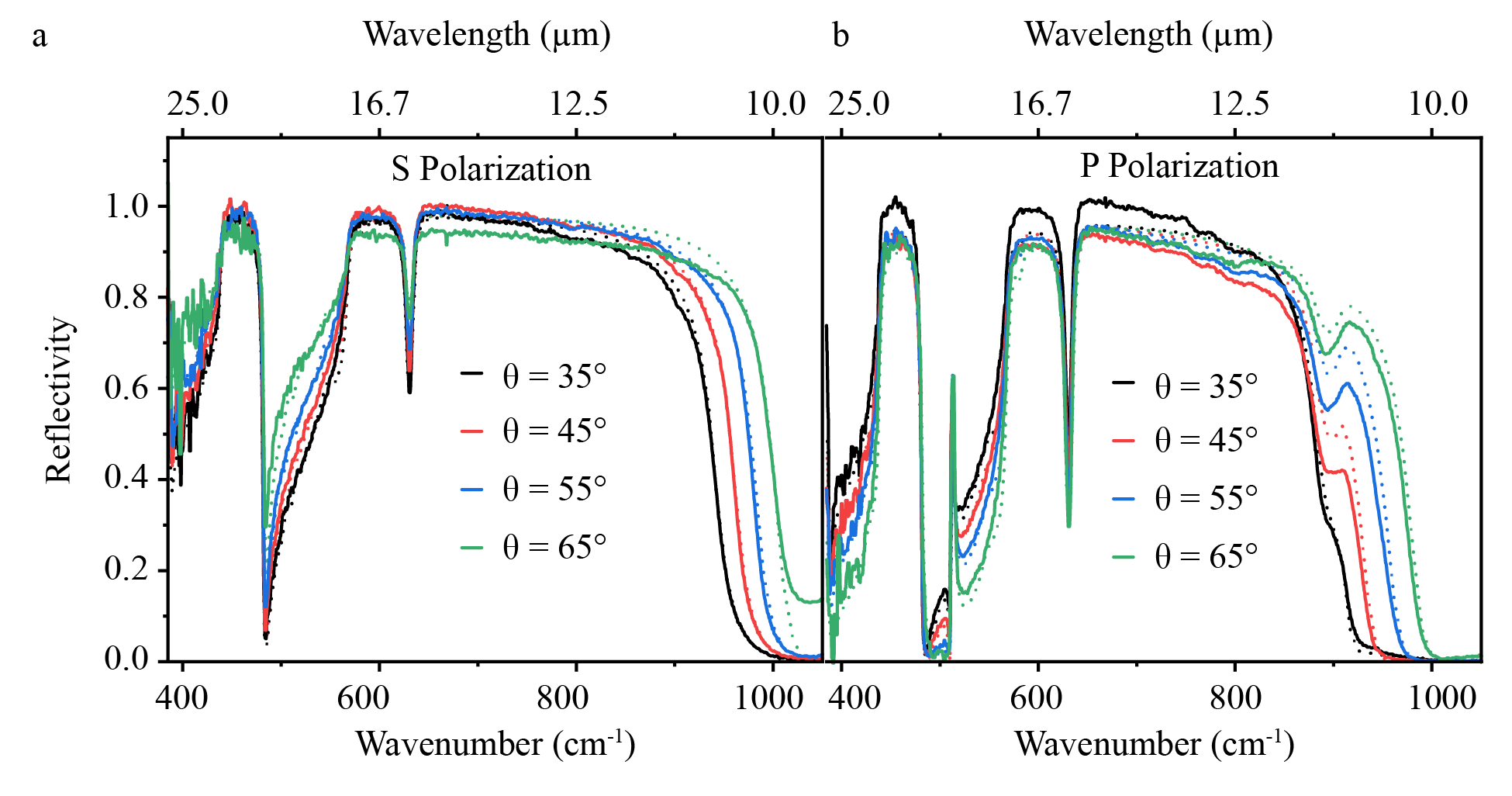}
  \caption{Experimental and modeled reflection spectra for bulk sapphire across a range of incident angles from $\theta = 35^\circ$ to $65^\circ$ for (a) $s$-polarized and (b) $p$-polarized excitations. For each angle of incidence, solid plots represent the experimental results, while the corresponding modeled spectra are depicted by dotted lines.
}
  \label{FigS2}
\end{figure}

A numerical analysis in the infrared (IR) range was performed using finite element method (FEM) in COMSOL Multiphysics. As shown in the left panel of Fig.~2c in the main text, a periodic unit cell with infinite periodic boundary conditions (BC) was simulated, using the nanostructure dimensions obtained from SEM images of the sample. The unit cell was excited by a plane wave with different polarizations (as shown in the left panel of Fig.~\ref{FigS1}) with various incident angles. The far-field reflection spectra for both unpatterened and structured $\ce{Al2O3}$ were investigated. The numerical results for the bulk material shows excellent agreement with the experimentally measured reflectivity data, as shown in Fig.~\ref{FigS2}, thereby confirming the accuracy for the permittivity values employed in the model \cite{schubert2000infrared}. 

\begin{figure}[H]
  \centering
\includegraphics[width=6.5 in, height=3.5 in]{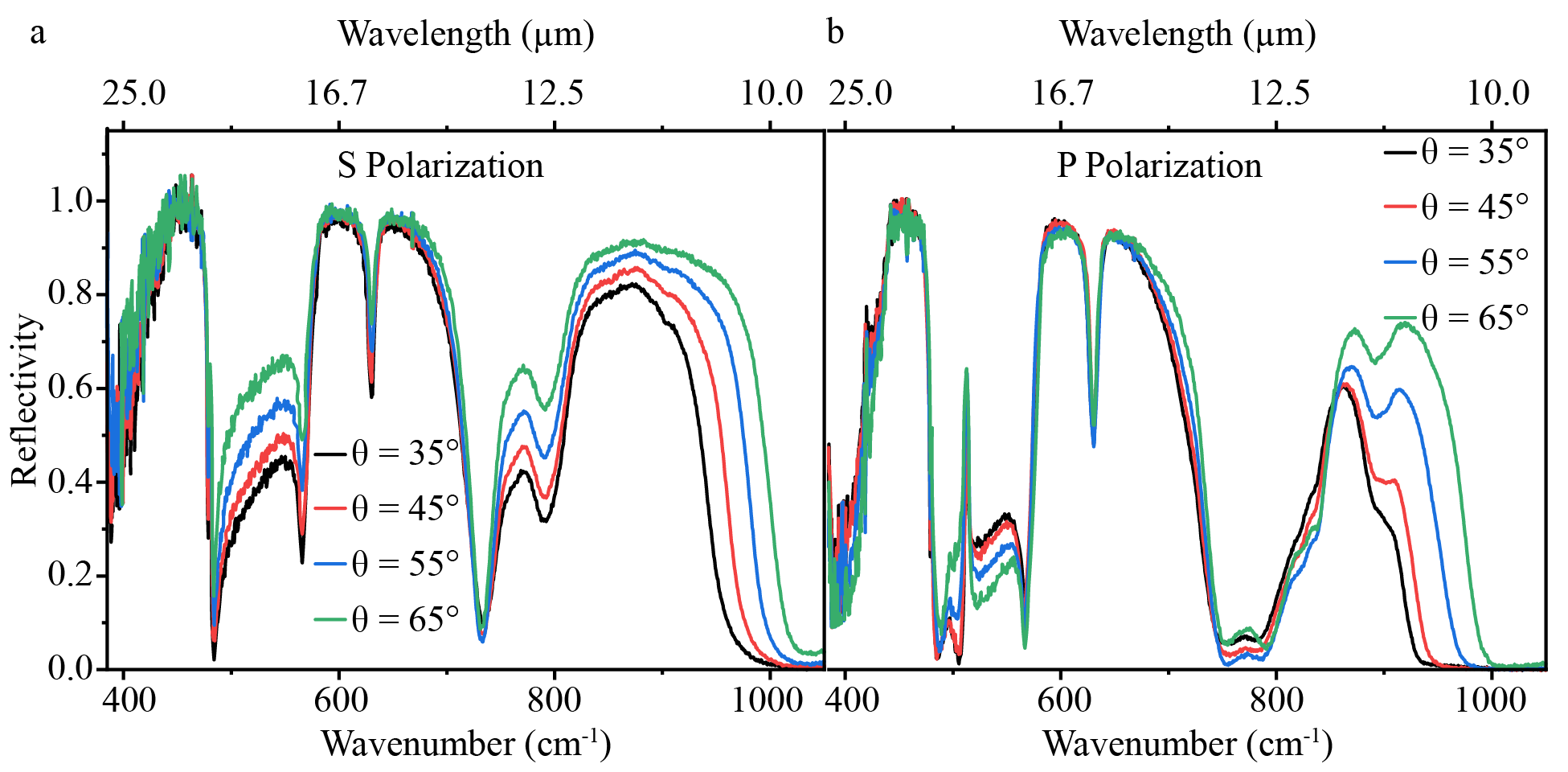}
  \caption{Experimental reflection spectra for sapphire NCA across a range of incident angles from $\theta = 35^\circ$ to $65^\circ$ for (a) $s$-polarized and (b) $p$-polarized excitations.
}
  \label{FigS3}
\end{figure}

Additionally, a comprehensive study of the nanostructure's reflectivity under various polarizations and angles of incidence confirms the presence of polariton modes in the NCA structure, as shown in Fig.~\ref{FigS3}. Furthermore, the structured system mimic the behavior of the unpatterned material in the peak strength near $\omega \approx 511~\text{cm}^{-1}$
and in the reflection band cut-off at high energies, as illustrated in Fig.~\ref{FigS3}. A comparison between the measured and simulated reflection spectra for NCA sapphire at an incident angle of $\theta = 45^\circ$ is presented in Figs.~ 2(a), 2(b), and 2(d) in the main text.

\begin{figure}[H]
  \centering
\includegraphics[width=3.5 in, height=3.5 in]{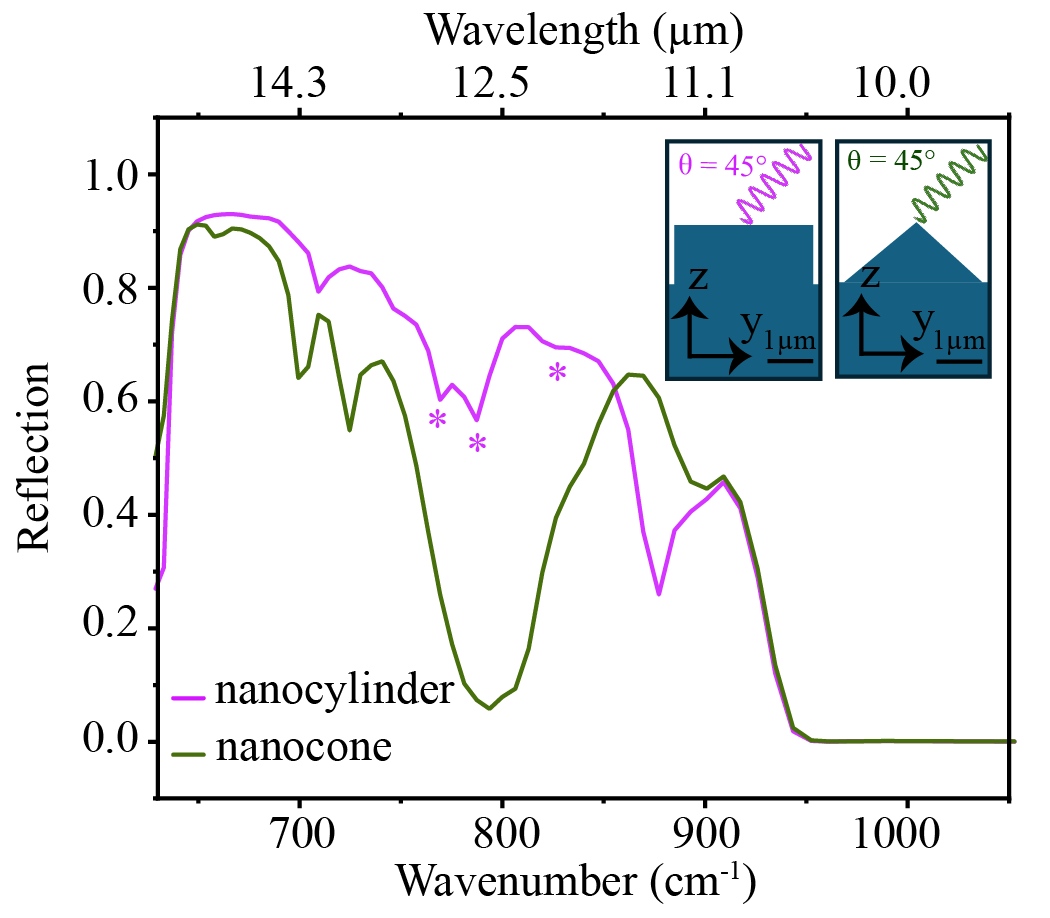}
  \caption{$p$-polarized calculated reflectivity corresponding to a fictitious nanocylinder structure with similar height and diameter compared with the NCA. The insets show a cross-sectional view of modeled unit cell.
}
  \label{FigS4}
\end{figure}

In an effort to uncover the impact of gradient width variation of the nanocones along the $z$ direction on mode~III, a parallel calculation was conducted for non-tapered sapphire nanocylinders with similar height of $h \simeq 2~\mu\text{m}$ and diameter of $d \simeq 3~\mu\text{m}$ as the nanocones. The results are shown in Figure~\ref{FigS4}, where three resonances are observed at $\omega \simeq 769~\text{cm}^{-1}$, $\omega \simeq 787~\text{cm}^{-1}$, and $\omega \simeq 817~\text{cm}^{-1}$ for the nanocylinders. These resonances fall within the bandwidth of mode~III, which spans from $\omega \simeq 719~\text{cm}^{-1}$ to $\omega \simeq 825~\text{cm}^{-1}$. This comparison reveals that tapering the nanocylinders along the $z$ direction to form nanocones allows for the simultaneous excitation of multiple SPhP modes at different frequencies, which can overlap and result in a broader overall resonance bandwidth \cite{cui2012ultrabroadband, yang2018broadband}.

\begin{figure}[H]
  \centering
\includegraphics[width=5.5 in, height=5.5 in]{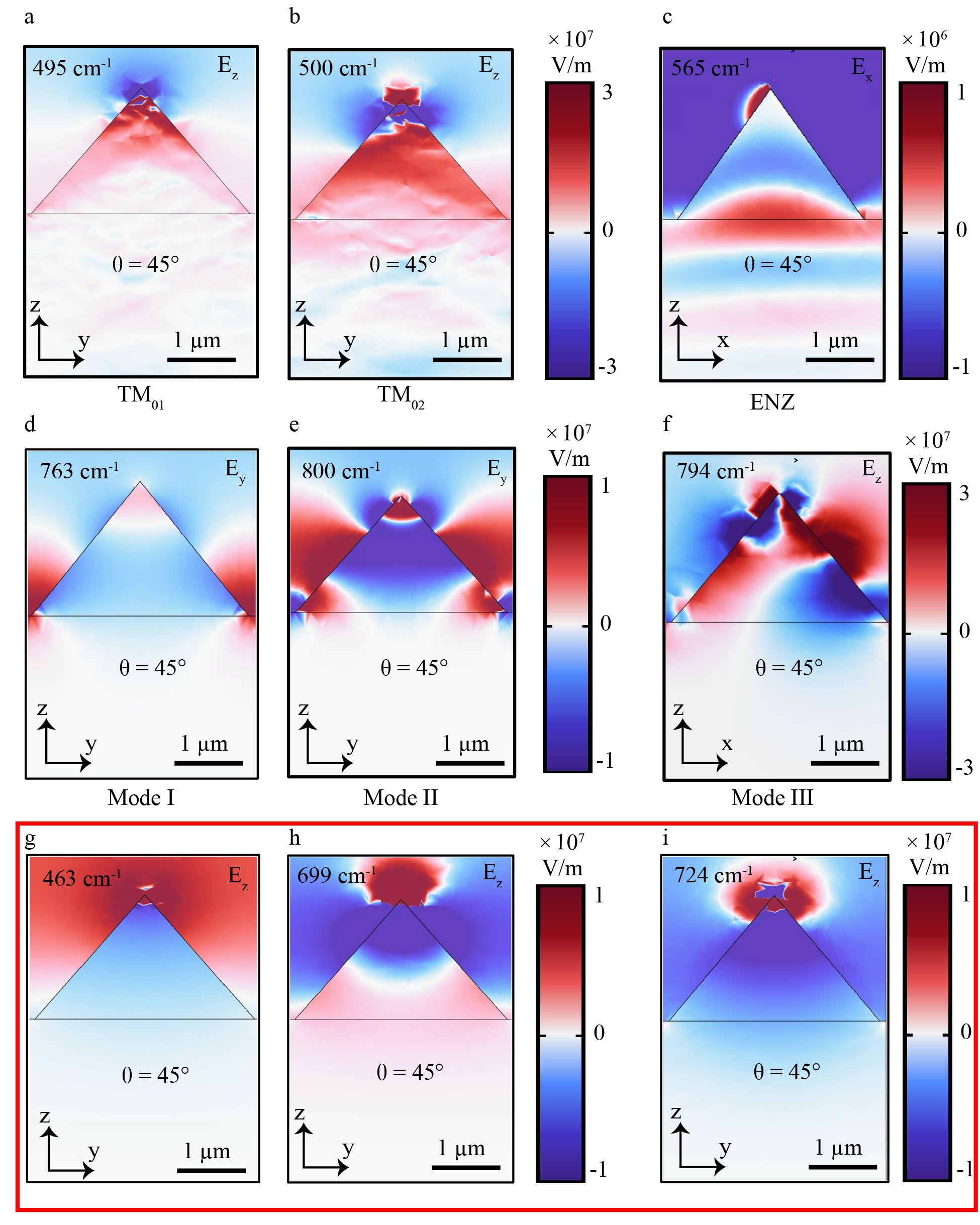}
  \caption{\textbf{E-field spatial distribution in sapphire NC for oblique incidence of $\theta = 45^\circ$}. $E_z$ simulated cross-sectional plots for HVPhPs at (a) 495~cm$^{-1}$, and (b) 500~cm$^{-1}$. $E_y$ simulated cross-sectional plot for (c) ENZ mode at 565~cm$^{-1}$, and calculated $E_y$ for $s$-polarized SPhP resonances at (d) 763~cm$^{-1}$, and (e) 800~cm$^{-1}$. $E_z$ surface plot for $p$-polarized SPhP at (f) 794~cm$^{-1}$, and for extra observed modes in the simulation at (g) 463~cm$^{-1}$, (h) 699~cm$^{-1}$, and (i) 724~cm$^{-1}$.
}
  \label{FigS5}
\end{figure}

Furthermore, the $E$-field spatial distribution corresponding to various exciting modes in NCA system for oblique incidence are shown in Fig.~\ref{FigS5}. 

\begin{figure}[H]
  \centering
\includegraphics[width=6.5 in, height=3.5 in]{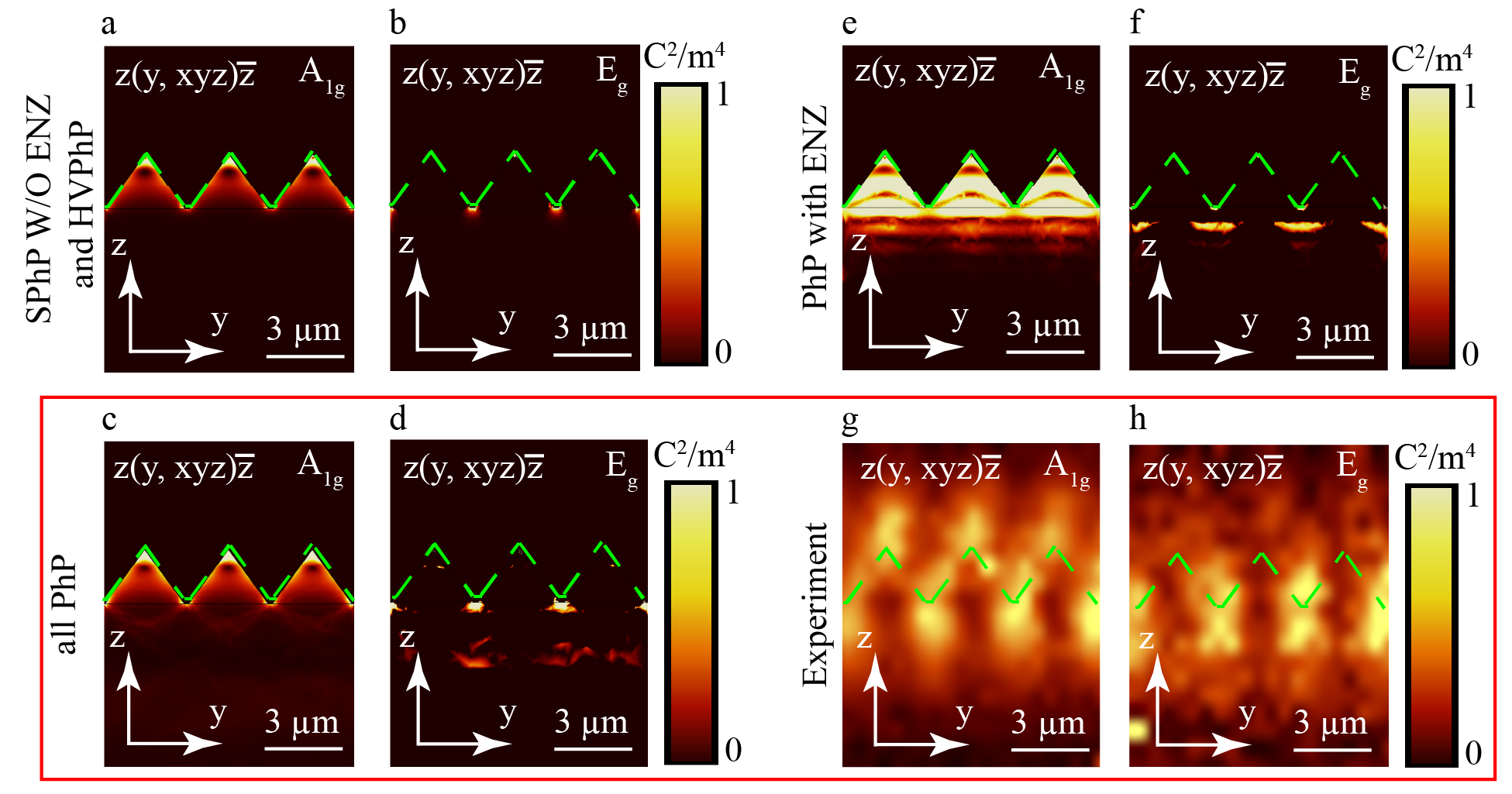}
  \caption{\textbf{Comparison between simulated and measured spatial Raman mappings in the ZY plane for three adjacent NCs:}.Calculated (a) $A_{1g}$ and (b) $E_g$ Raman intensity by considering SPhP modes. Calculated (c) $A_{1g}$ and (d) $E_g$ Raman intensity by considering all PhP modes including SPhP and HVPhP. Calculated (e) $A_{1g}$ and (f) $E_g$ phonon dispersion by considering PhPs and ENZ mode. Measured (g) $A_{1g}$ and (h) $E_g$ spatial Raman intensity.
}
  \label{FigS6}
\end{figure}

 To model the spatial distribution of Raman intensity (Figs.~4d, 4e, and Fig.~\ref{FigS6}), polarizability selection rules were applied to phonon polariton (PhP) eigenmodes of the NC structure~\cite{arledge2024mapping, bergeron2023probing}. The eigenmodes obtained from reflection spectra were subjected to the Raman selection rules for $A_{1g}$ and $E_g$ phonon symmetries. Finally, the Raman scattering intensity was reconstructed by summing over the selected polarization intensities of the PhP eigenmodes.
In an effort to investigate phonon dispersion, spatial Raman intensities for phonons were calculated using COMSOL Multiphysics for various fictitious cases. These cases include the coupling of $A_{1g}$ and $E_g$ phonons with surface phonon polaritons (SPhPs), all phonon polaritons (including SPhPs and hyperbolic volume phonon polaritons (HVPhPs)), and PhPs along with epsilon-near-zero (ENZ) mode. As shown in Figs.~\ref{FigS6}e and \ref{FigS6}f, the phonon dispersions arising from coupling $A_{1g}$ and $E_g$ phonons with both optical and ENZ modes deviate significantly from the experimental data. This discrepancy arises from the fundamentally different nature of ENZ and optical phonon modes. However, as demonstrated in Figs.~\ref{FigS6}c, \ref{FigS6}d, \ref{FigS6}g, and \ref{FigS6}h, excellent agreement is observed between the experimental data and the calculated spatial Raman mappings when all PhPs are coupled with the phonons in the NCA structure.
\label{sec: SI}
\newpage
\bibliography{ACSreferences}

\end{document}